\documentclass[preprint,aps]{revtex4}

\usepackage{graphicx}
\usepackage{dcolumn}
\usepackage{bm}

\def\ä{\"{a}}
\def\ü{\"{u}}
\def\ö{\"{o}}
\def\Ä{\"{A}}
\def\Ü{\"{U}}
\def\Ö{\"{O}}

\begin{document}

\title[]{Investigating Atomic Details of the CaF$_2$(111) Surface with a qPlus Sensor}

\author{Franz J. Giessibl}

\affiliation{\ Universit\ät Augsburg, Experimentalphysik 6, EKM,
Institut f\ür Physik, 86135 Augsburg, Germany}

\author{Michael Reichling}

\affiliation{\ Fachbereich Physik, Universit\ät Osnabr\ück,
Barbarastra{\ss}e 7, 49076 Osnabr\ück, Germany}

\begin{abstract}
The (111) surface of CaF$_2$ has been intensively studied with
large-amplitude frequency-modulation atomic force microscopy and
atomic contrast formation is now well understood. It has been
shown that the apparent contrast patterns obtained with a polar
tip strongly depend on the tip terminating ion and three
sub-lattices of anions and cations can be imaged. Here, we study
the details of atomic contrast formation on CaF$_2$(111) with
small-amplitude force microscopy utilizing the qPlus sensor that
has been shown to provide utmost resolution at high scanning
stability. Step edges resulting from cleaving crystals in-situ in
the ultra-high vacuum appear as very sharp structures and on flat
terraces, the atomic corrugation is seen in high clarity even for
large area scans. The atomic structure is also not lost when
scanning across triple layer step edges. High resolution scans of
small surface areas yield contrast features of anion- and cation
sub-lattices with unprecedented resolution. These contrast
patterns are related to previously reported theoretical results.

Submitted to Nanotechnology (Proceedings of NCAFM 2004)
\end{abstract}




\maketitle

\section{Introduction}
CaF$_2$ is an important material for science and technology, e.g.
as a lens material for 157\,nm lithography \cite{Letz:2004} or as
a high-bandgap insulating layer with almost perfect lattice match
for epitaxial insulating layers on silicon \cite{Smith:1984}. A
detailed knowledge of its surface structure and defects is
important. While thin CaF$_2$ layers can be imaged by scanning
tunneling microscopy \cite{Avouris:1989}, an atomic force
microscope (AFM) \cite{Binnig:1986b} is required for imaging
thicker layers \cite{Klust:2004} or bulk materials
\cite{Reichling:1999}. Crystalline CaF$_2$ has a face-centered
cubic lattice (see Fig. 1). The natural cleavage planes are the
$\{111\}$ planes \cite{Tasker:1979}, the corresponding surface
layers are trigonal arrangements of F$^-$-ions spaced by
$a_0/\sqrt{2} = 386.2$\,pm. Electrostatic energy considerations
lead to the conclusion that the CaF$_2$ (111) surface must be
terminated by a complete triple layer F$^-$-Ca$^{++}$-F$^-$ with a
F$^-$-layer at the surface \cite{Tasker:1979}.
\begin{figure}[h]
  \includegraphics[width=8cm]{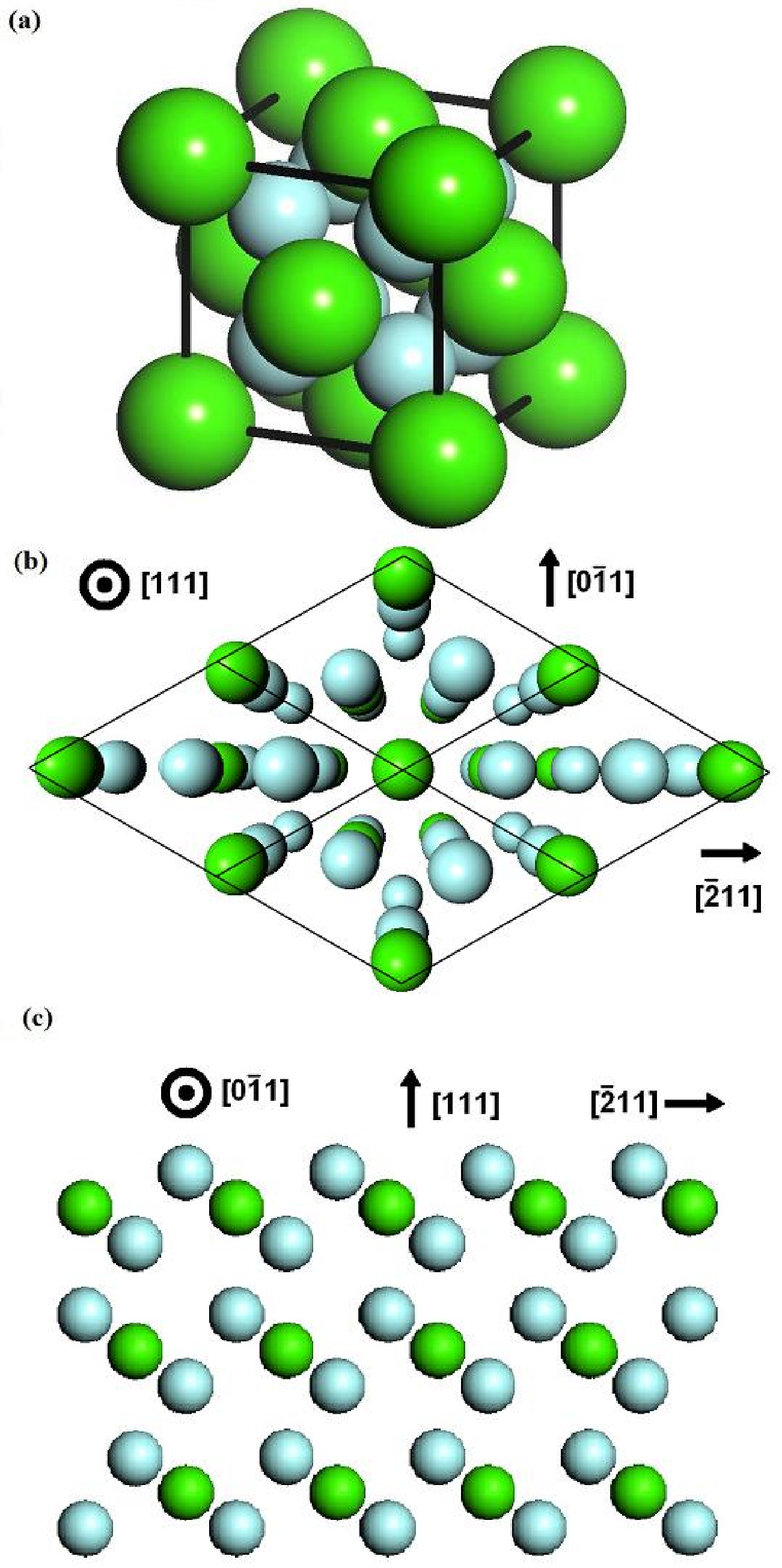}
  \caption{Crystal structure of CaF$_2$,
forming a face-centered cubic lattice with a lattice constant of
$a_0 = 546.2$\,pm. The Ca$^{++}$-ions are represented by green
spheres, the F$^{-}$-ions by blue spheres. The basis of the
lattice consists of three atoms with a Ca$^{++}$-ion located at
the origin of the fcc-lattice and two F$^{-}$-ions located at
$(x,y,z)=\pm(a_0/4,a_0/4,a_0/4)$ with an interionic distance of
$\sqrt{3}/4$ $a_0 = 237$\,pm. (a) Perspective view of the cubic
unit cell. (b) Top view of a CaF$_2$ (111) surface with a surface
layer consisting of F$^{-}$-ions. The surface layer is a trigonal
arrangement of F$^-$-ions spaced by $a_0/\sqrt{2} = 386.2$\,pm,
the angle between the surface lattice vectors is $60 ^\circ$. (c)
View parallel to the surface along the [$\underline{1}10$]
direction, showing the electrically neutral triple layers of
F$^-$-Ca$^{++}$-F$^-$ ions spaced by $a_0/\sqrt{3} = 315.35$\,pm.
Because the surface should be terminated by complete triple layers
(see text), step heights should be integer multiples of
315.35\,pm. The Ca$^+$-layer is 79\,pm below the surface
F$^-$-layer, and the second F$^-$-layer is 158\,pm lower than the
surface layer.}
  \label{Fig1}
\end{figure}
In contrast to the (001) cleavage planes of alkali-halides, the
CaF$_2$ (111) surface offers a reference sample where the atomic
contrast in experimental images is tightly connected to the
signature of the electric charge on the tip \cite{Foster:2002}.
AFM studies of the CaF$_2$ surface are available on the atomic
scale \cite{Reichling:1999,Barth:2001b,Reichling:2002bk} as well
as on a larger scale \cite{Schick:2004}. While a consistent
understanding of the atomic contrast has been achieved, step
structures have so far proven hard to be imaged with traditional
large-amplitude AFM at atomic resolution, and theoretical
considerations that predict a change in contrast pattern when
imaging in the repulsive regime \cite{Foster:2002bk,Foster:2002}
have so far not been verified experimentally. The use of small
amplitudes helps to attenuate the disturbing long-range
interaction forces \cite{Giessibl:2003} and enables
non-destructive atomic imaging in the repulsive regime
\cite{Giessibl:2001a}.  The qPlus sensor can be operated with high
stability at sub-nm amplitudes \cite{Giessibl:2004}, motivating to
revisit CaF$_2$ (111).

\section{Challenge of long range forces on steps}
In an ideal (hypothetical) AFM, the probe would consist of a
single atom. In reality, the front atom needs to be supported by
other atoms, forming a mesoscopic tip that is connected to a
cantilever. The tip-sample forces in AFM are a sum of long- and
short range contributions. While chemical bonding forces between
single atoms decay exponentially with increasing distance,
van-der-Waals (vdW)- and electrostatic forces decay only at an
inverse power law and therefore have a much longer range. While
chemical bonding forces for two atoms at close distance can be
much greater than the vdW forces between two atoms, the total vdW
force between tip and sample is typically significantly greater
than the chemical bonding force between front atom and sample. The
long-range contribution is often modelled by a vdW interaction of
a spherical tip with radius $R$, yielding a long-range force given
by
\begin{equation}\label{FvdW}
   F_{vdW}(z) = -\frac{A_H R}{6z^2}
\end{equation}
\cite{Israelachvili:1991} where $A_H$ is the Hamaker constant.
In frequency modulation AFM \cite{Albrecht:1991}, the tip-sample force is
not measured directly. Instead, the averaged gradient of the the tip-sample force
with respect to the surface normal leads to a frequency shift $\Delta f$ of a
cantilever with an unperturbed
eigenfrequency $f_0$, spring constant $k$ and oscillation amplitude $A$.
A vdW-force as in Eq. \ref{FvdW} leads to a normalized frequency shift
$\gamma = (\Delta f/f_0) k A^{3/2}$ given by
\begin{equation}\label{dflr}
   \gamma_{vdW}(z) = -A_H R \frac{A^{3/2}}{6(z^2+2Az)^{3/2}}
\end{equation}
\cite{Hoelscher:1999}. The long-range force is proportional to the
tip radius $R$, and thus sharp tips minimize the long-range
contribution to the normalized frequency shift. This long-range
force is a challenge for high-resolution AFM, in particular when
scanning across steps where the long-range force changes as a
function of the lateral sample position. Guggisberg et al.
\cite{Guggisberg:2000b} have investigated the frequency shift
difference on upper and lower terraces (FREDUL) in Si(111) and
found a voltage dependent variation corresponding to
5\,fNm$^{0.5}$ when compensating the contact potential difference
and 30\,fNm$^{0.5}$ for a bias voltage of 2\,V. As short range
contributions to $\gamma$ are on the order of 1\,fNm$^{0.5}$, this
clearly stresses the challenges faced when attempting AFM at
atomic resolution across steps. While sporadic reports of atomic
resolution across step edges by large amplitude FM-AFM have been
reported (e.g. \cite{Bennewitz:2001,Allers:2002bk}), it requires a
cantilever with an exceptionally sharp tip. Here we show that when
using small amplitudes, atomic resolution across step edges is
even possible with relatively blunt tips.

\begin{figure}[h]
  \includegraphics[width=10cm]{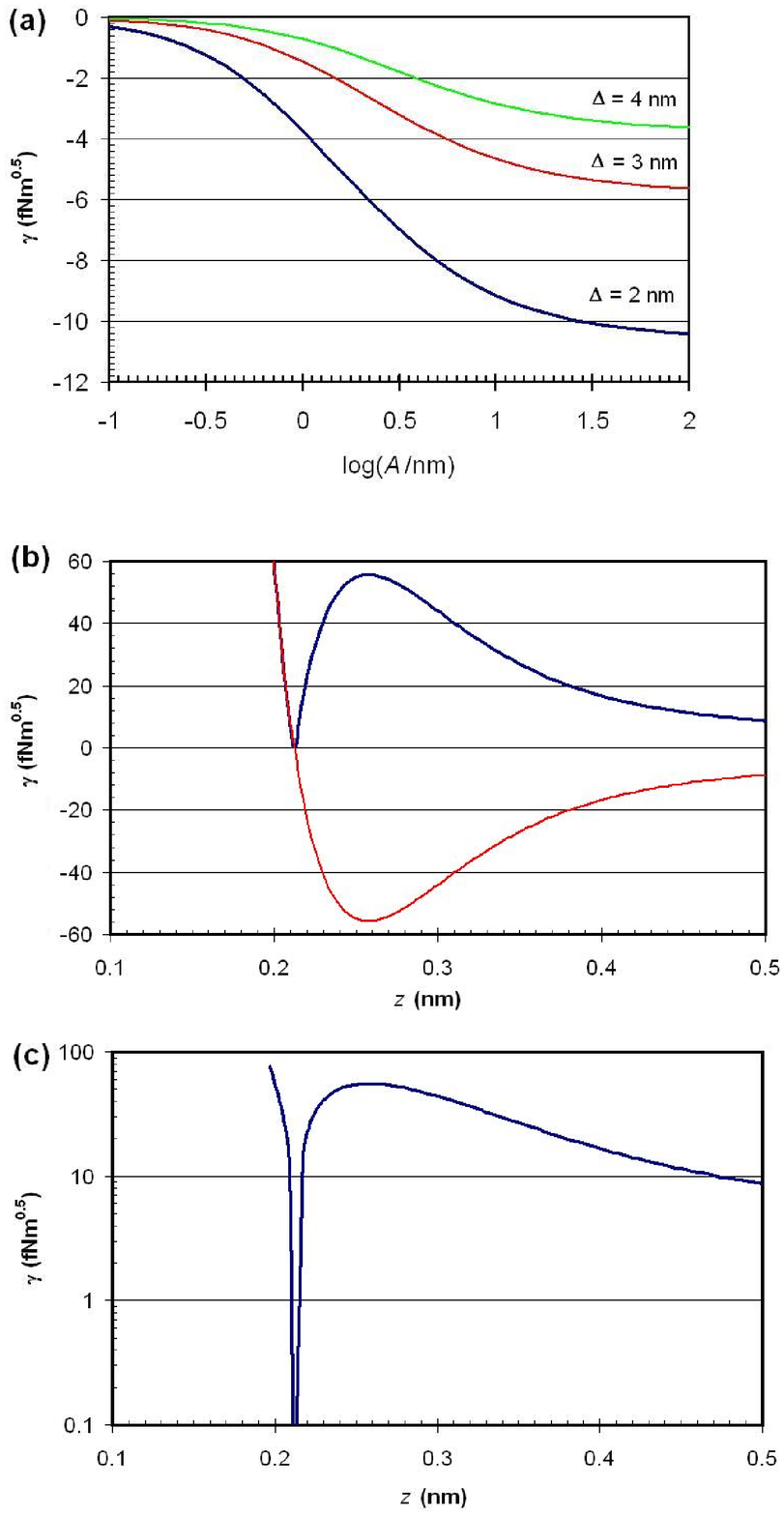}
  \caption{(a) Simulated normalized frequency shift $\gamma$ as
a function of tip cluster height $\Delta$ and amplitude $A$. (b)
Simulated normalized frequency shift $\gamma$ and $|\gamma|$ as a
function of distance $z$ for an amplitude of $A=1$\,nm and the
tip-sample forces described in Eqs. \ref{FvdW} and \ref{Vts} with
$A_H=1$\,eV, $R=100$\,nm, $\Delta = 1$\,nm, $F_0=12.15$\,nN,
$\kappa = 16.3$\,nm$^{-1}$. (c) Logarithmic display of $|\gamma|$,
showing that a logarithmic filter provides a feedback signal that
is roughly linear with $z$ for $z$-values outside of $\approx
0.21$\,nm.}
  \label{Fig2}
\end{figure}

In the beginning of an AFM experiment, the tip often collects a
cluster consisting of sample material with a height $\Delta$,
which also reduces the long-range force. Figure \ref{Fig2} (a)
shows the calculated dependence of the long-range component of
$\gamma$ as a function of tip cluster height $\Delta$ and
amplitude $A$ after Eq. 2.
It is clearly evident that the long-range contribution of $\gamma$
is greatly reduced when using small amplitudes and having a high
tip cluster.

For short range forces caused by covalent bonding, a Morse
potential
\begin{equation}\label{Vts}
   V_{ts}(z) = E_{bond}(-2\exp^{-\kappa (z-\sigma)}+\exp^{-2\kappa (z-\sigma)})
\end{equation}
is a fair approximation for the short-range part of the tip-sample
interaction \cite{Perez:1998}, where $E_{bond}$ is the bonding
energy, $\kappa$ is the inverse interaction range and $\sigma$ is
the equilibrium distance. When imaging ionic crystals, the
dominant short-range force is electrostatic in origin
\cite{Foster:2002bk}, but due to the periodic arrangement of
positive and negative charges, the net electrostatic force decays
roughly exponential with a decay constant of $\kappa=2 \pi/a_S$
\cite{Giessibl:1992a} where $a_S$ is the length of the surface
unit vector. If the tip atom is not a single charged atom, but a
cluster of the sample material, the decay rate of the
electrostatic force is estimated at $\kappa=4 \pi/a_S$. The
short-range part of the normalized frequency shift is given by
\begin{equation}\label{dfsr}
   \gamma_{sr}(z) = F_0\sqrt{1/\kappa}(-2 \exp^{-\kappa (z-\sigma)}+\sqrt{2} \exp^{-2\kappa (z-\sigma)})
\end{equation}
where $F_0$ is the tip-sample force at $z=\sigma$ for amplitudes
larger than $1/\kappa$ \cite{Giessibl:1997b}. It is noted, that
this model for the short-range force is only qualitative but
serves well to discuss the challenges of atomic imaging. A
detailed study of the short-range forces in AFM on CaF$_2$ can be
found in references \cite{Foster:2002bk,Foster:2002}. Figure 2 (b)
shows the dependence of the normalized frequency shift $\gamma(z)$
for a tip-sample force composed of long- and short-range
contributions as described by Eqs. \ref{FvdW} and \ref{Vts}.

Rather than
using $\Delta f$ directly as a feedback signal, in our experiment we routed $\Delta
f$ through a rectifier and a logarithmic filter before using it
for feedback. Rectifying $\Delta f$ prevents catastrophic tip
crashes due to inadvertent jumps into the repulsive imaging regime
as previously described by Ueyama et al. \cite{Ueyama:1998}. The
use of a logarithmic filter improves tracking on steps and other
sharply inclined topography features, because it provides a feedback
signal that is more linear with distance (see Fig. 2 (c)).
A disadvantage of rectifying $\Delta f$ is that for small magnitudes of
$\gamma$ ($|\gamma|<55$\,fNm$^{0.5}$ in Fig. 2 (c)), a one-to-one relation between
$\gamma$ and $z$ is not present and two $z$ values with $\partial\gamma/\partial z<0$ exist,
thus the $z$-feedback can find two $z$-values $z_i$ where the stable-feedback-conditions
 $\gamma(z_i)=\gamma_{setpoint}$ and $\partial\gamma/\partial z<0$ for a distance regime $z = z_i+\epsilon$. Figure
\ref{Fig2} (b) and (c) show that for a setpoint of $|\gamma| =
20$\,fNm$^{0.5}$, two distances are possible ($z_1\approx 0.2$\,nm
and $z_2\approx 0.4$\,nm). The $z$-intervals [$
z_1-\epsilon$,$z_1+\epsilon$] and [$ z_2-\epsilon$,$z_2+\epsilon$]
where stable operation is possible can have a width $2\epsilon$
reaching a few hundred pm. In some experiments we experienced
jumps in topographic data where stable topographic imaging was
possible for two $z$-values separated by approximately 0.2\,nm. A
switch from $z_2$ to $z_1$ can be triggered intentionally when
scanning rapidly across a rising step, while a reverse switch is
triggered by scanning across a falling step (not shown here).

\begin{figure}[h]
  \includegraphics[width=8cm]{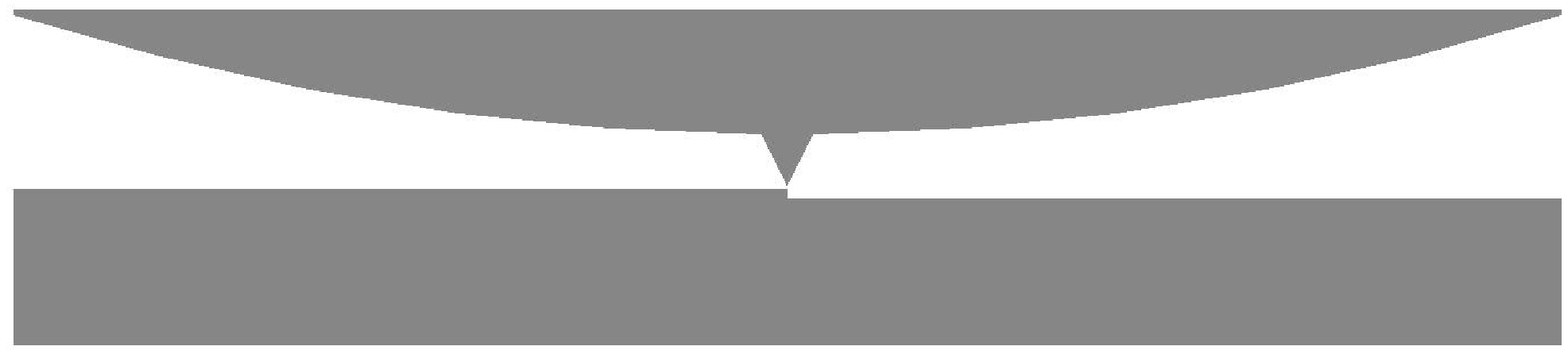}
  \caption{Schematic true-to-scale representation of the lower section of a spherical tip with a radius of
  100\,nm and a micro-asperity with a height of 2\,nm close to a sample with a 315\,pm step.}
  \label{Fig3}
\end{figure}

Figure \ref{Fig3} is a schematic view of a tip over a step edge.
The attractive interaction is greater over the lower terrace than
over the higher terrace. This causes a challenge when attempting
to image step edges with atomic resolution. As noted above, a
reduction of the disturbing long-range force can be achieved by
using small oscillation amplitudes, high tip clusters and sharp
tips.

\section{Experimental}
The sample used in this study was a CaF$_2$ crystal with a size of
approximately $2\times 5 \times 10$\,mm$^3$ (Karl Korth, Kiel,
Germany), glued onto a $11\times 14 $\,mm$^2$ large sample holder
plate. The sample was cleaved along a predetermined breaking line
in situ in the (111) plane at a pressure of $5\times 10^{-7}$\,Pa
and within one minute transferred to the microscope where it was
kept a pressure of $5\times 10^{-8}$\,Pa. As CaF$_2$ (111) is not
very reactive, we obtained atomic resolution on this single cleave
until five days after cleaving and collected approximately 4000
images within that period.

The microscope (AutoProbe VP by Park Scientific Instruments,
Sunnyvale, USA) \cite{Giessibl:1994a} was modified for qPlus
sensor operation. The force sensor is a standard qPlus sensor
\cite{Giessibl:2000b} with a base frequency of $f_0 = 16740$\,Hz,
a $Q$-factor of 1700 and a stiffness of $k=1800$\,N/m. An etched
tungsten tip with an estimated tip radius of 100\,nm was used as a
tip. Frequency-to-voltage conversion was done with a commercial
phase-locked-loop detector (EasyPLL by Nanosurf AG, Liestal,
Switzerland). Images were recorded in the topographic mode at
constant frequency shift, supplemented by constant-height
measurements of small sample sections that allow us to estimate
the long- and short-range contributions. Drift correction of the
acquired data was performed using a commercial software package
(SPIP Scanning Probe Image Processor Version 2.21 by Image
Metrology, Lyngby, Denmark).

\section{Results}

We started the scan by slowly decreasing the setpoint of $\Delta
f$ while monitoring the contrast. When performing AFM on CaF$_2$
with soft cantilevers ($k\approx 10$\,N/m), surface charges
usually cause unstable feedback conditions and often the crystal
is heated for a few hours to remove these charges. Surprisingly,
we did not have to heat the sample for obtaining stable imaging.
Whether this is caused by the large stiffness of the qPlus sensor
or the metallic tip is not yet determined. A large area scan shown
in Fig. 4 shows flat terraces, separated by steps with heights of
integer multiples (1--4) of 320\,pm $\pm$ 15\,pm, in excellent
agreement with the expected triple-layer height of 315\,pm.
\begin{figure}[h]
  \includegraphics[width=14cm]{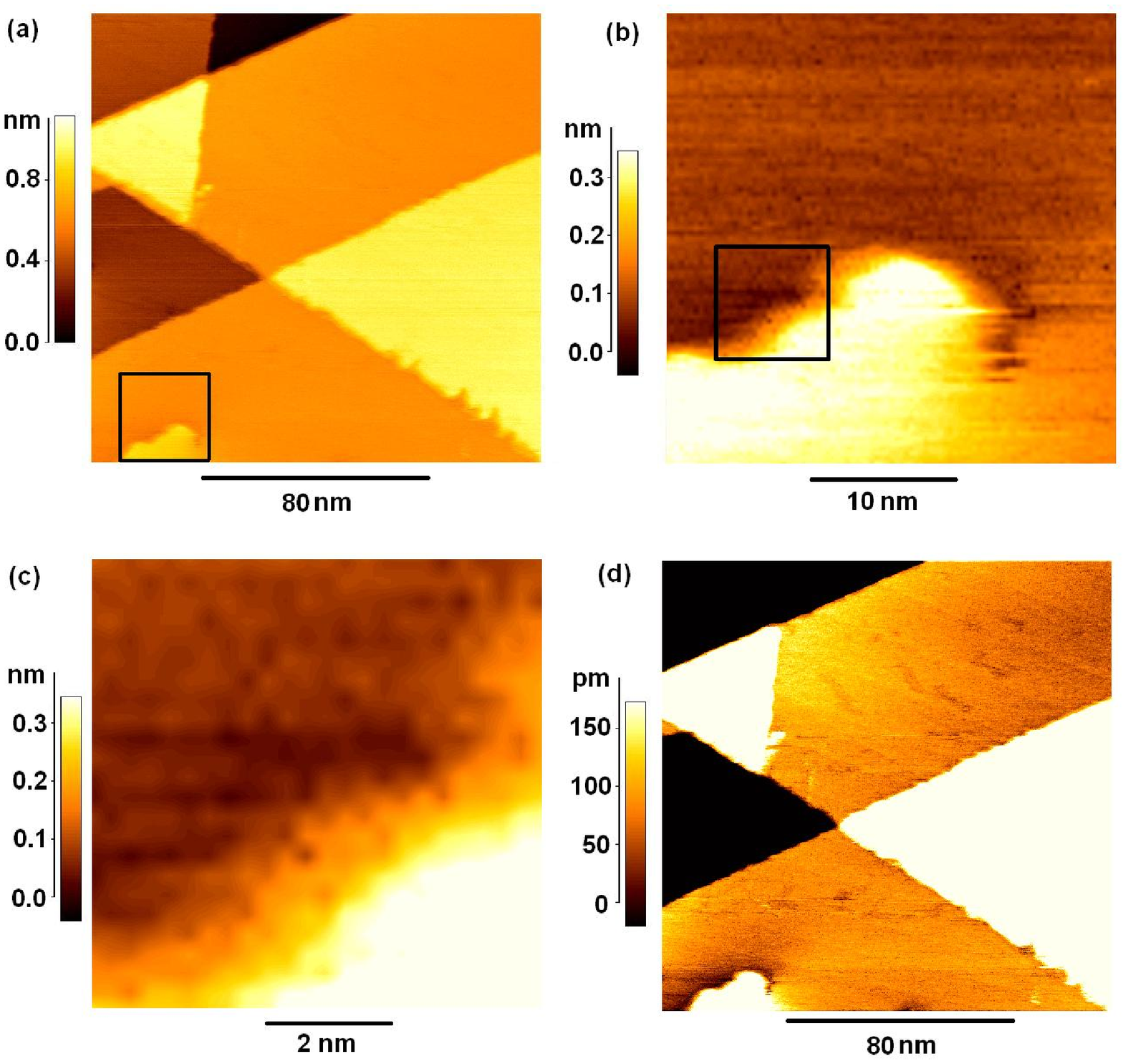}
  \caption{Overview AFM image and magnified views of a CaF$_2$ (111) surface.
  Imaging parameters: Scanning speed 96\,nm/s, $A=625$\,pm, $\Delta f=+7.32$\,Hz, $\gamma=+12.3$\,fNm$^{0.5}$.
  (a) Image size $160 \times 160$\,nm$^2$, with four different terrace heights
  spaced by integer multiples (1--4) of 315\,pm. (b) Image size $28.85 \times 28.85$\,nm$^2$,
  showing a screw dislocation with a height of 315\,pm. (c) Zoom into
  the step at the screw dislocations, image size $8 \times 8$\,nm$^2$ showing atomic contrast.
  (d) Same image as (a) with 5-fold increased contrast, showing patched and linear structures on single terraces.}
  \label{Fig4}
\end{figure}
The data shown in Fig. \ref{Fig4} is taken from a single topographic measurement with an
image size of $160 \times 160$\,nm$^2$ (512 by 512 pixels). Figure \ref{Fig4} (a)
shows the full image, (b) and (c) are magnifications of areas indicated by the frames.
Figures \ref{Fig4} (b) and (c) show the center of a screw dislocation. The step height at the left
edge is 314\,pm. The image was recorded with a positive frequency shift, i.e. at repulsive forces.
Figure \ref{Fig4} (d) shows the same data as (a) with five-fold $z$-contrast. The small
patches on otherwise flat terraces are probably caused by local surface- or sub-surface charges.

\begin{figure}[h]
  \includegraphics[width=8cm]{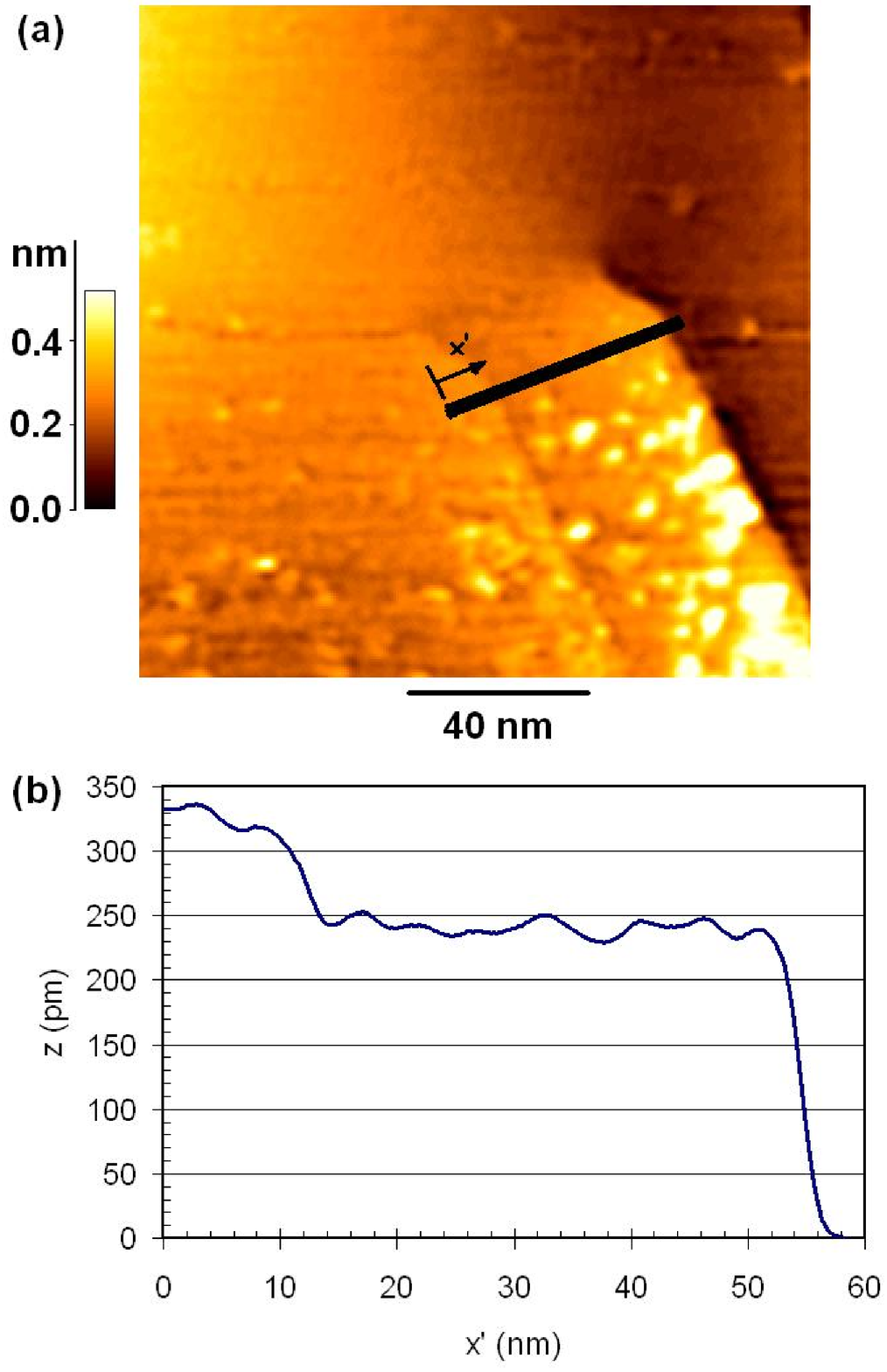}
  \caption{AFM image of a screw dislocation on a CaF$_2$ (111) surface.
  Imaging parameters: Scanning speed 400\,nm/s, image size
  $150 \times 150$\,nm$^2$, $A=1.25$\,nm, $\Delta f=-0.74$\,Hz, $\gamma=-3.5$\,fNm$^{0.5}$.}
  \label{Fig5}
\end{figure}

Figure \ref{Fig5} (a) shows the center of a different screw
dislocation located 1.22\,$\mu$m to the left of the screw
dislocation shown in Fig. 4 (b). The profile in Fig. \ref{Fig5}
(b) shows that the step height is initially only approximately
240\,pm, followed by another step with a height of only 80\,pm.
The unusual step heights cannot be explained by the commonly
accepted crystallography of CaF$_2$ (111). A reduced step height
of 240\,pm could possibly be explained by the long-range force
contributions explained in the text describing Fig. 3 (further
below, we find an experimental step height of only 275\,pm for a
step imaged in the attractive mode, see Fig. 8). However, a step
with a height of 80\,nm could not be explained by such an effect.
Other explanations, like a double tip effect, appear unlikely
because of the large lateral step distance of more than 40\,nm.

A possibly obvious explanation appears to contradict Taskers
theorem \cite{Tasker:1979}: a region with a width of 40\,nm could
be stripped of the F$^-$ ions. In equilibrium, a (111) face
exposing Ca$^{++}$ ions is forbidden. On the other hand, cleaving
a crystal is not an equilibrium process, and our crystal was not
annealed after cleaving (Figure 5 (a) was taken about 30\,min
after cleaving). Furthermore, the large strain fields in the
vicinity of a screw dislocation may help to violate the
charge-neutrality principle. The large number of adsorbants on the
40\,nm wide odd-stacked terrace points to a highly reactive
surface region, possibly caused by a large surface charge density.
We note, that Ca$^{++}$ terminated surfaces of CaF$_2$ (111) have
not been reported in the literature so far to our knowledge, but
neither have atomic images of screw dislocations. So far, we have
seen this unusual step height in one out of two screw dislocation
centers (incidentally, the second screw dislocation center shown
in Fig. 4 was recorded 5 days after cleaving), and further studies
to elucidate the dislocation morphology of CaF$_2$ are planned.

\begin{figure}[h]
  \includegraphics[width=14cm]{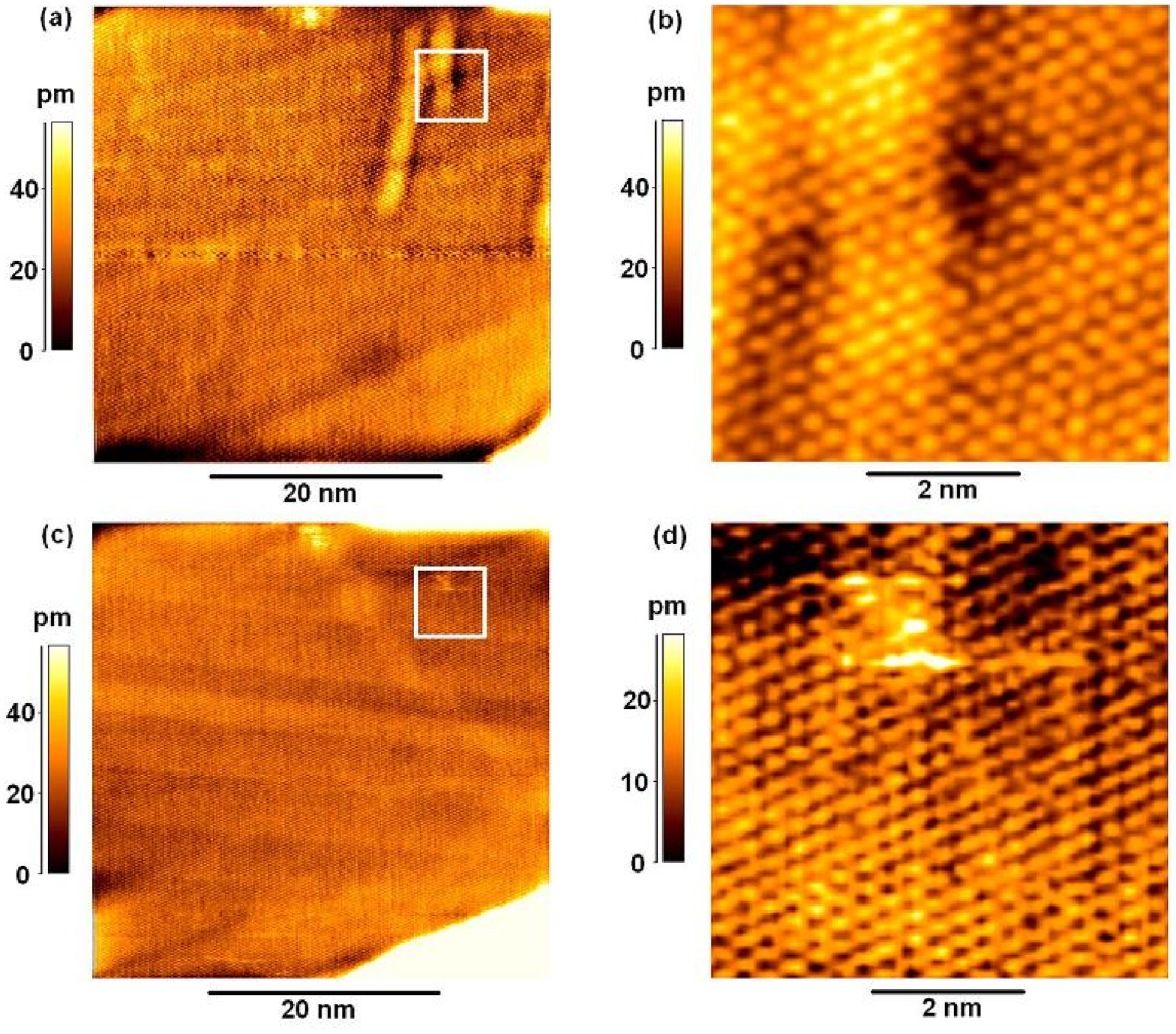}
  \caption{High-resolution topographic images of CaF$_2$ (111).
  Imaging parameters: Scanning speed 20\,nm/s, image size
   $40 \times 40$\,nm$^2$ (a,c), $6.1 \times 6.1$\,nm$^2$ (b,d), $A=625$\,pm, $\Delta f=+3.66$\,Hz, $\gamma=6.15$\,fNm$^{0.5}$.
   Acquistion time for (a) and (c) is 34\,min, (c) was taken right after (a).}
  \label{Fig6}
\end{figure}

Figure \ref{Fig6} (a) shows a large scan showing approximately
12,000 atoms. There is a double triple-layer step in the right
bottom corner, but otherwise the terrace is flat. The $\mu$-shaped
structure is possibly caused by surface charges. Repp et al.
\cite{Repp:2004} have recently shown that charged Au-adatoms on NaCl
can cause a significant relaxation of the underlying NaCl lattice that remains
stable even when electrons tunnel through this adatom. Discharging of
this adatom and thus switching its state is only possible when a voltage pulse
with sufficient pulse height is applied. From Repp et al.'s experiment,
it is conceivable that charged in-plane surface atoms on insulators may
also be stabilized by lattice
distortions that could cause slight deviations of an otherwise flat sample.

The
magnified view in Fig. \ref{Fig6} (b) shows a structural defect,
and Fig. \ref{Fig6} (c) images the same area after 34\,min.
The
$\mu$-shaped structure has disappeared in \ref{Fig6} (c), but the
magnified view in \ref{Fig6} (d) shows that the structural defect
is still there. The fast scan
(0.25 lines/s, starting at the bottom in (a) and at the top in (c)) was
horizontal, the slow scan vertical. Thus, the step at the right bottom in
Fig. 6 (c) was imaged 68 min after the step in Fig. 6 (a). Because of
thermal drift, the steps appear shifted in (a) and (c), and the magnified
view of the structural defect also shifts between Figs. 6 (b) and (d).

\begin{figure}[h]
  \includegraphics[width=14cm]{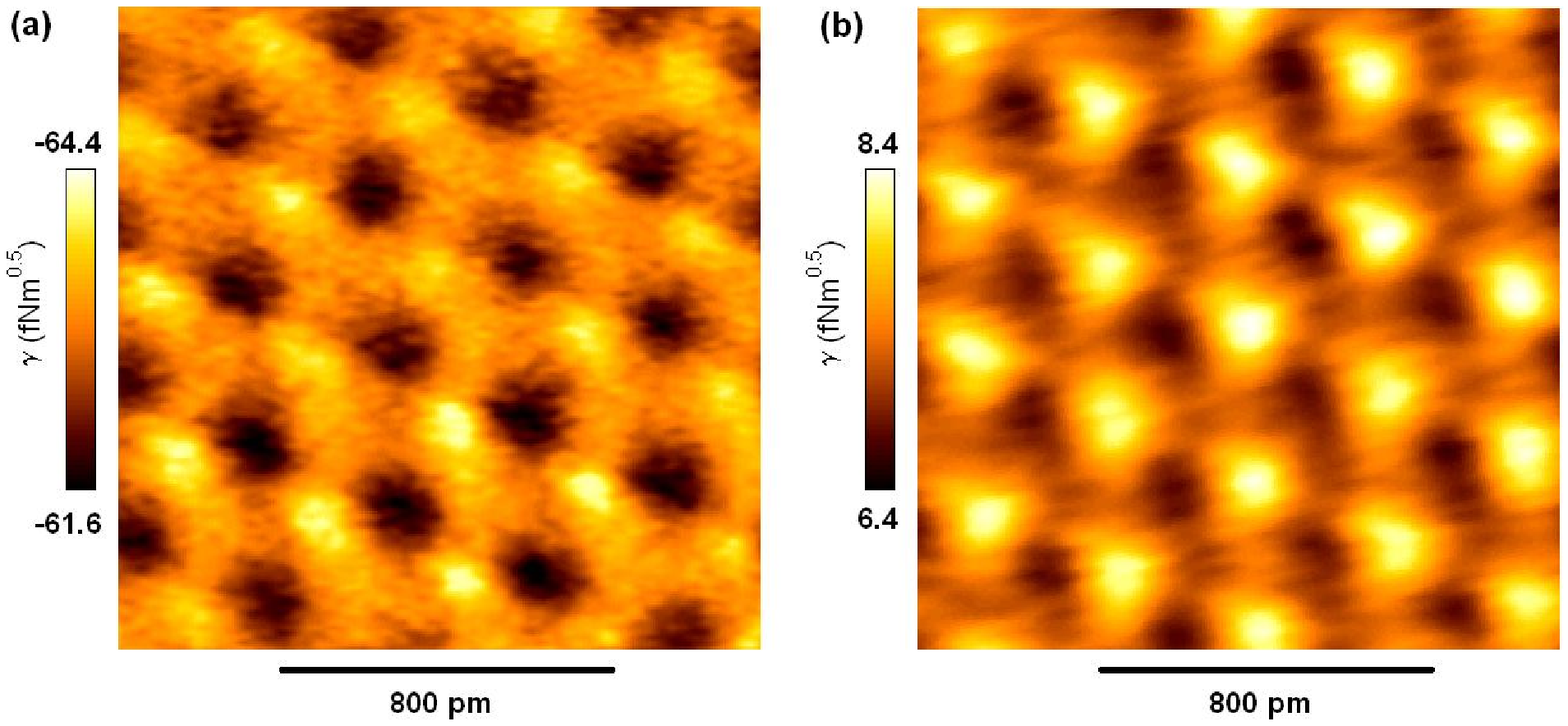}
  \caption{High-resolution constant height images of CaF$_2$ (111) in the attractive (a) and repulsive mode (b).
  Imaging parameters: Image size $1.36 \times 1.36$\,nm$^2$,
  (a) scanning speed 4\,nm/s, $A=1.25$\,nm, $\Delta f= -13.3 \pm 0.3$\,Hz, $\gamma=-63 \pm 1.4$\,fNm$^{0.5}$
  (b) scanning speed 16\,nm/s, $A=625$\,pm, $\Delta f= +4.4 \pm 0.3$\,Hz, $\gamma=+7.4\pm 1 $\,fNm$^{0.5}$.}
  \label{Fig7}
\end{figure}

Figure \ref{Fig7} (a) is a high-resolution image imaged in the
attractive mode which, according to previous calculations
\cite{Foster:2002bk,Foster:2002}, is produced by a positively
terminated tip. The orientation of the maxima, minima and saddle
points in the image shows that the sample is oriented as indicated
in Fig. 1 (b) and (c). The maxima are attributed to the surface
F$^-$-layer, the minima to the Ca$^{++}$-layer that are 79\,pm
lower and the saddle points to the second F$^-$-layer that is
158\,pm lower than the surface layer (see Fig. 1b). Figure
\ref{Fig7} (b) shows the same area imaged in a repulsive mode. As
predicted by Foster et al. \cite{Foster:2002bk}, the contrast
changes -- while the absolute minima in Fig. 7 (a) are adjacent to
the right of the absolute maxima, they are left of the absolute
maxima in the repulsive data shown in Fig. 7 (b). The magnitude of
the $\gamma$-contrast in Fig. 7 (a) of $\pm 1.4$\,fNm$^{0.5}$
agrees very well with the calculated value of $\approx \pm
1$\,fNm$^{0.5}$ (Foster et al. find a contrast of up to 8\,Hz
using a cantilever with $k=6$\,N/m, $f_0=84$\,kHz and $A=23$\,nm,
see p. 327 and 333 in \cite{Foster:2002bk}). The absolute value
for $\gamma$ according to theory is $-38.6$\,fNm$^{0.5}$ (see p.
328 in \cite{Foster:2002bk}), while we find almost twice that
value in Fig. 7 (b). This deviation is most likely due to the
fairly large radius of our tip. In large-amplitude FM-AFM,
experimental values for $\gamma$ range from approximately
$-250$\,fNm$^{0.5}$ \cite{Reichling:1999} to $-85$\,fNm$^{0.5}$
\cite{Klust:2004}. In the small amplitude experiments presented
here, we observed atomic resolution in the attractive mode in a
range from $-100$\,fNm$^{0.5}$ to $-25$\,fNm$^{0.5}$. The
agreement of the parameter range for $\gamma$ in large- and small
amplitude regimes, where the basic imaging parameters $k,A,f_0$
and $\Delta f$ differ by orders of magnitude underlines the
validity of $\gamma$ as a universal figure describing tip-sample
interaction in FM-AFM.

\begin{figure}[h]
  \includegraphics[width=12cm]{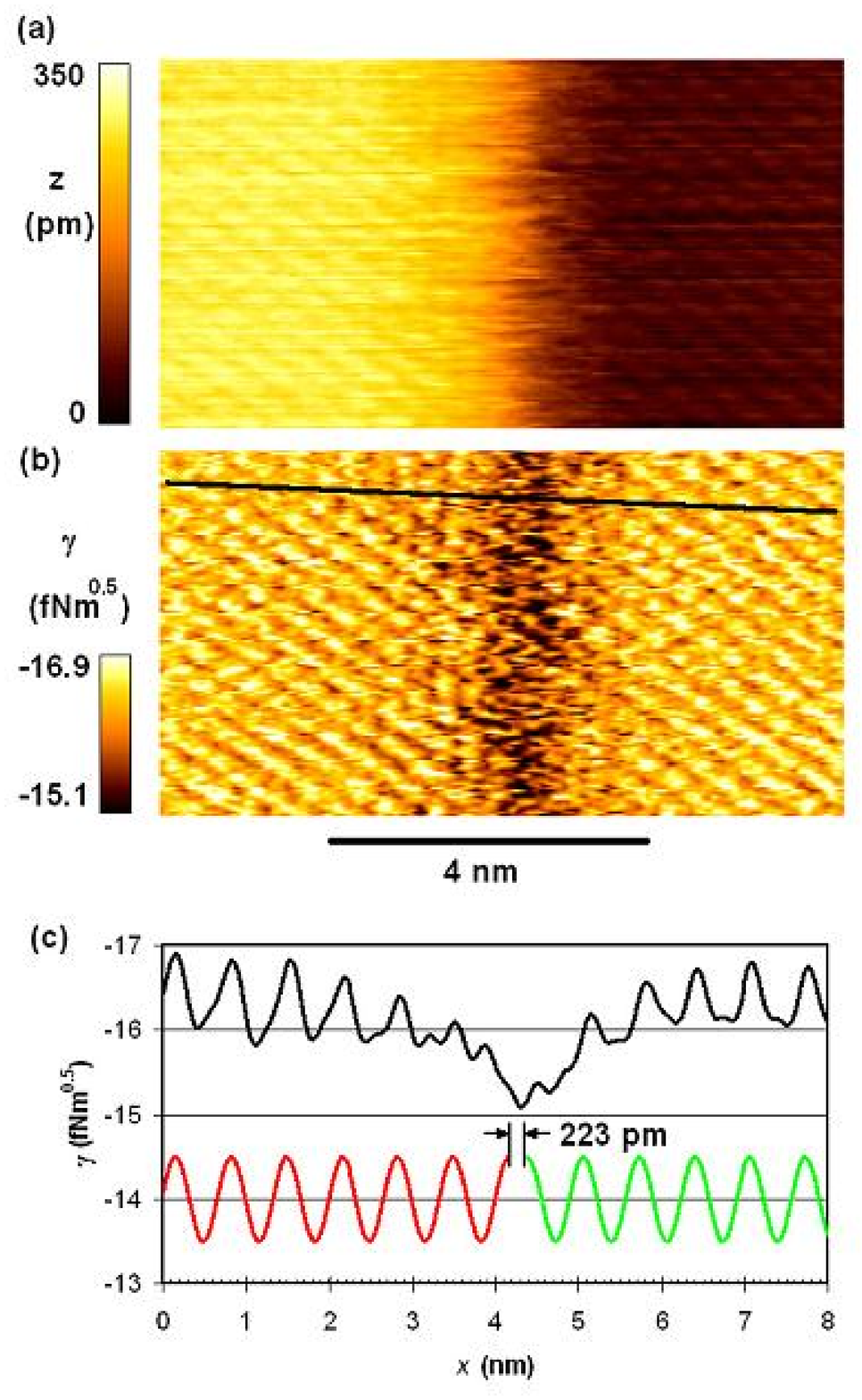}
  \caption{High-resolution topographic image (a) and frequency shift image (b) of CaF$_2$ (111) across a single
  step. The measured step height is 275\,pm, 13\% smaller than the
  expected step height of 315\,pm (see text).
  Imaging parameters: Scanning speed 8.7\,nm/s, image size
  $8.6 \times 4.5$\,nm$^2$, $A=625$\,pm, $\Delta f=-9.5$\,Hz, $\gamma=-16 \pm 0.9$\,fNm$^{0.5}$.
  (c) Contour line along the black trace indicated in (b). A filtered version of (b) was used to produce the contour.}
  \label{Fig8}
\end{figure}

Figure \ref{Fig8} (a) is a topographic image across a monostep with a height of 315\,pm recorded in the
attractive mode. Due to the lateral variation of the vdW force, the apparent step height is only 275\,pm, but
atomic resolution is present on both the higher and the lower terrace. The setpoint of the normalized frequency
shift was $\gamma=-16$\,fNm$^{0.5}$ in this image, but due to finite feedback speed, the actual frequency shift
has shown small variations as shown in Fig. 8 (b). The line analysis in Fig. 8 (c) of the error signal along the black line in
Fig. 8 (b) shows that the F$^-$ ions are shifted by approximately 223\,pm to the right as expected from Fig. 1 (c).

Figure 9 demonstrates that the front atom of the tip is not a metal atom from the original tip but consists of an ion
or ionic cluster picked up from the surface.
The top section shows contrast as expected from a negative tip termination, and the bottom shows inverted
contrast as if the charge of the front atom is inverted. Such a contrast change could be caused by a
CaF$_2$ tip cluster that is shifted or flipped during the scan, exposing a F$^-$ ion initially and a Ca$^{++}$ tip ion in the lower
section of the image.

\begin{figure}[h]
  \includegraphics[width=14cm]{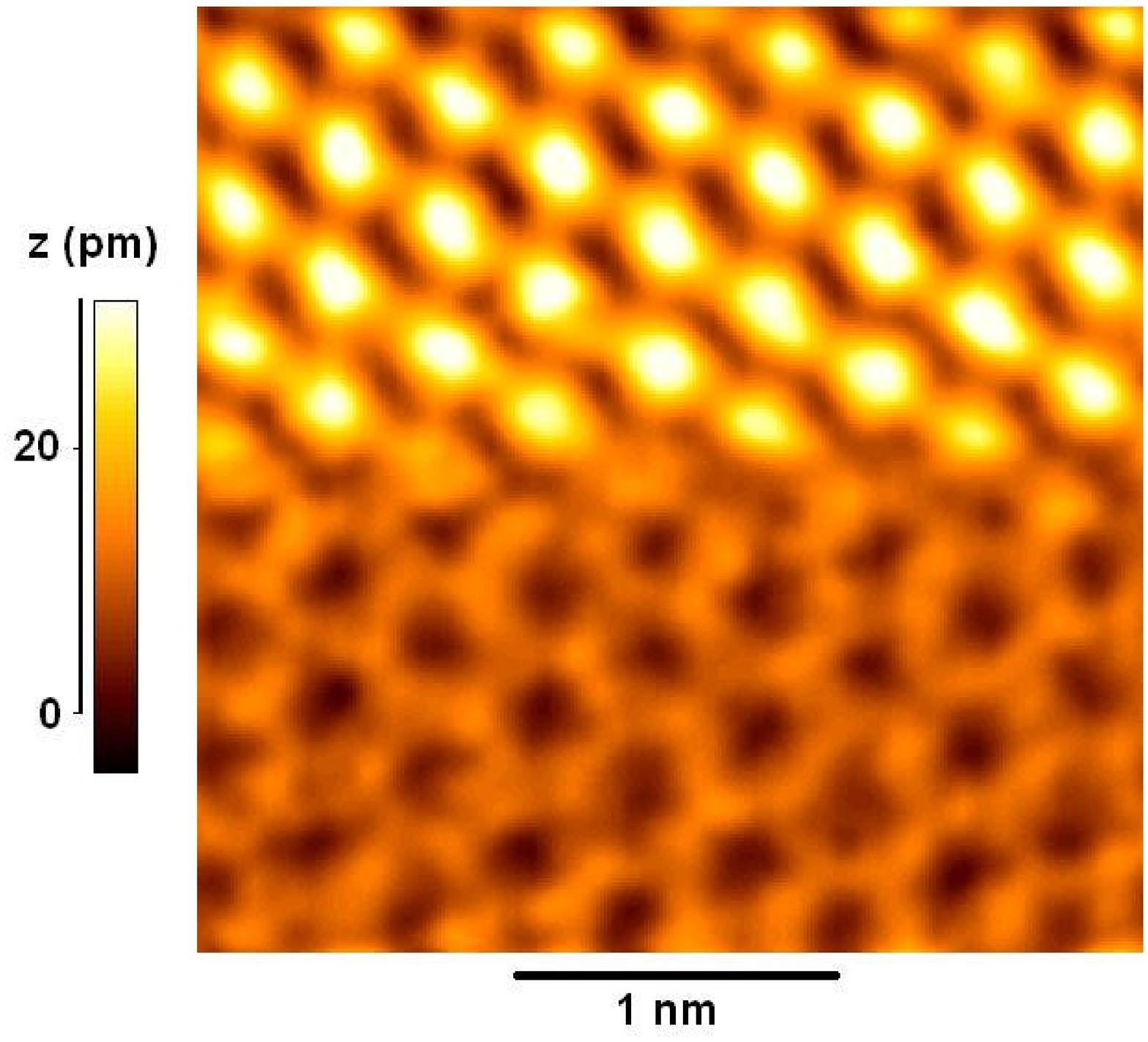}
  \caption{Topographic image of CaF$_2$ (111) during which a tip change occurs.
  Imaging parameters: Scanning speed 4.8\,nm/s, image size
  $2.9 \times 2.9$\,nm$^2$,  $A=625$\,pm, $\Delta f=-8.8$\,Hz, $\gamma=-14.8$\,fNm$^{0.5}$.}
  \label{Fig9}
\end{figure}

\section{Conclusion and Summary}

In conclusion, we have shown that the use of stiff cantilevers such as the qPlus sensor operated with
 small amplitudes even allows the imaging of steps on CaF$_2$ (111) with atomic resolution both in attractive and
 repulsive mode. Large terraces with up to 12,000 atoms have been imaged.
 The theoretical prediction about a shift in contrast when switching from attractive to repulsive imaging \cite{Foster:2002bk,Foster:2002}
 has been verified. The centers of screw dislocations have been imaged for the first time, and we found that
 small sample areas can exist that are not terminated by complete triple layers. Further improvements are expected when
 using qPlus sensors with sharp tips \cite{Giessibl:2001a}, because that would allow the further attenuation of the disturbing long-range contributions.

\begin{acknowledgments}
This work is supported by the Bundesministerium f\"{u}r Bildung
und Forschung (project EKM13N6918). We thank Ch. Schiller for the
preparation of the qPlus sensor used in this study and J. Mannhart for support.
M.R. gratefully acknowledges support from the Deutsche Forschungsgemeinschaft.
\end{acknowledgments}

\bibliography{2004fjg}

\begin{thebibliography}{28}
\expandafter\ifx\csname natexlab\endcsname\relax\def\natexlab#1{#1}\fi
\expandafter\ifx\csname bibnamefont\endcsname\relax
  \def\bibnamefont#1{#1}\fi
\expandafter\ifx\csname bibfnamefont\endcsname\relax
  \def\bibfnamefont#1{#1}\fi
\expandafter\ifx\csname citenamefont\endcsname\relax
  \def\citenamefont#1{#1}\fi
\expandafter\ifx\csname url\endcsname\relax
  \def\url#1{\texttt{#1}}\fi
\expandafter\ifx\csname urlprefix\endcsname\relax\def\urlprefix{URL }\fi
\providecommand{\bibinfo}[2]{#2}
\providecommand{\eprint}[2][]{\url{#2}}

\bibitem[{\citenamefont{Letz}(2004)}]{Letz:2004}
\bibinfo{author}{\bibfnamefont{M.}~\bibnamefont{Letz}},
  \bibinfo{journal}{{}Physik Journal} \textbf{\bibinfo{volume}{3}},
  \bibinfo{pages}{43} (\bibinfo{year}{2004}).

\bibitem[{\citenamefont{Smith et~al.}(1984)\citenamefont{Smith, Phillips,
  Augustyniak, and Stiles}}]{Smith:1984}
\bibinfo{author}{\bibfnamefont{T.~P.} \bibnamefont{Smith}},
  \bibinfo{author}{\bibfnamefont{J.~M.} \bibnamefont{Phillips}},
  \bibinfo{author}{\bibfnamefont{W.~M.} \bibnamefont{Augustyniak}},
  \bibnamefont{and} \bibinfo{author}{\bibfnamefont{P.~J.}
  \bibnamefont{Stiles}}, \bibinfo{journal}{{}Appl. Phys. Lett.}
  \textbf{\bibinfo{volume}{45}}, \bibinfo{pages}{907} (\bibinfo{year}{1984}).

\bibitem[{\citenamefont{Avouris and Wolkow}(1989)}]{Avouris:1989}
\bibinfo{author}{\bibfnamefont{P.}~\bibnamefont{Avouris}} \bibnamefont{and}
  \bibinfo{author}{\bibfnamefont{R.}~\bibnamefont{Wolkow}},
  \bibinfo{journal}{{}Appl. Phys. Lett.} \textbf{\bibinfo{volume}{55}},
  \bibinfo{pages}{1074} (\bibinfo{year}{1989}).

\bibitem[{\citenamefont{Binnig et~al.}(1986)\citenamefont{Binnig, Quate, and
  Gerber}}]{Binnig:1986b}
\bibinfo{author}{\bibfnamefont{G.}~\bibnamefont{Binnig}},
  \bibinfo{author}{\bibfnamefont{C.~F.} \bibnamefont{Quate}}, \bibnamefont{and}
  \bibinfo{author}{\bibfnamefont{C.}~\bibnamefont{Gerber}},
  \bibinfo{journal}{Phys. Rev. Lett.} \textbf{\bibinfo{volume}{56}},
  \bibinfo{pages}{930} (\bibinfo{year}{1986}).

\bibitem[{\citenamefont{Klust et~al.}(2004)\citenamefont{Klust, Ohta, Bostwick,
  Yu, Ohuchi, and Olmstead}}]{Klust:2004}
\bibinfo{author}{\bibfnamefont{A.}~\bibnamefont{Klust}},
  \bibinfo{author}{\bibfnamefont{T.}~\bibnamefont{Ohta}},
  \bibinfo{author}{\bibfnamefont{A.~A.} \bibnamefont{Bostwick}},
  \bibinfo{author}{\bibfnamefont{Q.}~\bibnamefont{Yu}},
  \bibinfo{author}{\bibfnamefont{F.~S.} \bibnamefont{Ohuchi}},
  \bibnamefont{and} \bibinfo{author}{\bibfnamefont{M.~A.}
  \bibnamefont{Olmstead}}, \bibinfo{journal}{{}Phys. Rev. B}
  \textbf{\bibinfo{volume}{69}}, \bibinfo{pages}{035405}
  (\bibinfo{year}{2004}).

\bibitem[{\citenamefont{Reichling and Barth}(1999)}]{Reichling:1999}
\bibinfo{author}{\bibfnamefont{M.}~\bibnamefont{Reichling}} \bibnamefont{and}
  \bibinfo{author}{\bibfnamefont{C.}~\bibnamefont{Barth}},
  \bibinfo{journal}{Phys. Rev. Lett.} \textbf{\bibinfo{volume}{83}},
  \bibinfo{pages}{768} (\bibinfo{year}{1999}).

\bibitem[{\citenamefont{Tasker}(1979)}]{Tasker:1979}
\bibinfo{author}{\bibfnamefont{P.~W.} \bibnamefont{Tasker}},
  \bibinfo{journal}{J. Phys. C: Solid State Phys.}
  \textbf{\bibinfo{volume}{12}}, \bibinfo{pages}{4977} (\bibinfo{year}{1979}).

\bibitem[{\citenamefont{Foster et~al.}(2002{\natexlab{a}})\citenamefont{Foster,
  Barth, Shluger, Nieminen, and Reichling}}]{Foster:2002}
\bibinfo{author}{\bibfnamefont{A.~S.} \bibnamefont{Foster}},
  \bibinfo{author}{\bibfnamefont{C.}~\bibnamefont{Barth}},
  \bibinfo{author}{\bibfnamefont{A.~L.} \bibnamefont{Shluger}},
  \bibinfo{author}{\bibfnamefont{R.~M.} \bibnamefont{Nieminen}},
  \bibnamefont{and}
  \bibinfo{author}{\bibfnamefont{M.}~\bibnamefont{Reichling}},
  \bibinfo{journal}{Phys. Rev. B} \textbf{\bibinfo{volume}{66}},
  \bibinfo{pages}{235417} (\bibinfo{year}{2002}{\natexlab{a}}).

\bibitem[{\citenamefont{Barth et~al.}(2001)\citenamefont{Barth, Foster,
  Reichling, and Shluger}}]{Barth:2001b}
\bibinfo{author}{\bibfnamefont{C.}~\bibnamefont{Barth}},
  \bibinfo{author}{\bibfnamefont{A.}~\bibnamefont{Foster}},
  \bibinfo{author}{\bibfnamefont{M.}~\bibnamefont{Reichling}},
  \bibnamefont{and} \bibinfo{author}{\bibfnamefont{A.~L.}
  \bibnamefont{Shluger}}, \bibinfo{journal}{J. Phys.: Condens. Matter}
  \textbf{\bibinfo{volume}{13}}, \bibinfo{pages}{2061} (\bibinfo{year}{2001}).

\bibitem[{\citenamefont{Reichling and Barth}(2002)}]{Reichling:2002bk}
\bibinfo{author}{\bibfnamefont{M.}~\bibnamefont{Reichling}} \bibnamefont{and}
  \bibinfo{author}{\bibfnamefont{C.}~\bibnamefont{Barth}}, in
  \emph{\bibinfo{booktitle}{Noncontact Atomic Force Microscopy}}, edited by
  \bibinfo{editor}{\bibfnamefont{S.}~\bibnamefont{Morita}},
  \bibinfo{editor}{\bibfnamefont{R.}~\bibnamefont{Wiesendanger}},
  \bibnamefont{and} \bibinfo{editor}{\bibfnamefont{E.}~\bibnamefont{Meyer}}
  (\bibinfo{publisher}{Springer Berlin Heidelberg New York},
  \bibinfo{year}{2002}), chap.~\bibinfo{chapter}{6}, pp.
  \bibinfo{pages}{109--124}.

\bibitem[{\citenamefont{Schick et~al.}(2004)\citenamefont{Schick, Dabringhaus,
  and Wandelt}}]{Schick:2004}
\bibinfo{author}{\bibfnamefont{M.}~\bibnamefont{Schick}},
  \bibinfo{author}{\bibfnamefont{H.}~\bibnamefont{Dabringhaus}},
  \bibnamefont{and} \bibinfo{author}{\bibfnamefont{K.}~\bibnamefont{Wandelt}},
  \bibinfo{journal}{J. Phys.: Condens. Matter} \textbf{\bibinfo{volume}{16}},
  \bibinfo{pages}{L33} (\bibinfo{year}{2004}).

\bibitem[{\citenamefont{Foster et~al.}(2002{\natexlab{b}})\citenamefont{Foster,
  Shluger, Barth, and Reichling}}]{Foster:2002bk}
\bibinfo{author}{\bibfnamefont{A.}~\bibnamefont{Foster}},
  \bibinfo{author}{\bibfnamefont{A.}~\bibnamefont{Shluger}},
  \bibinfo{author}{\bibfnamefont{C.}~\bibnamefont{Barth}}, \bibnamefont{and}
  \bibinfo{author}{\bibfnamefont{M.}~\bibnamefont{Reichling}}, in
  \emph{\bibinfo{booktitle}{Noncontact Atomic Force Microscopy}}, edited by
  \bibinfo{editor}{\bibfnamefont{S.}~\bibnamefont{Morita}},
  \bibinfo{editor}{\bibfnamefont{R.}~\bibnamefont{Wiesendanger}},
  \bibnamefont{and} \bibinfo{editor}{\bibfnamefont{E.}~\bibnamefont{Meyer}}
  (\bibinfo{publisher}{Springer Berlin Heidelberg New York},
  \bibinfo{year}{2002}{\natexlab{b}}), chap.~\bibinfo{chapter}{17}, pp.
  \bibinfo{pages}{305--348}.

\bibitem[{\citenamefont{Giessibl}(2003)}]{Giessibl:2003}
\bibinfo{author}{\bibfnamefont{F.~J.} \bibnamefont{Giessibl}},
  \bibinfo{journal}{Rev. Mod. Phys.} \textbf{\bibinfo{volume}{75}},
  \bibinfo{pages}{949} (\bibinfo{year}{2003}).

\bibitem[{\citenamefont{Giessibl et~al.}(2001)\citenamefont{Giessibl,
  Hembacher, Bielefeldt, and Mannhart}}]{Giessibl:2001a}
\bibinfo{author}{\bibfnamefont{F.~J.} \bibnamefont{Giessibl}},
  \bibinfo{author}{\bibfnamefont{S.}~\bibnamefont{Hembacher}},
  \bibinfo{author}{\bibfnamefont{H.}~\bibnamefont{Bielefeldt}},
  \bibnamefont{and} \bibinfo{author}{\bibfnamefont{J.}~\bibnamefont{Mannhart}},
  \bibinfo{journal}{{}Applied Physics A} \textbf{\bibinfo{volume}{72}},
  \bibinfo{pages}{15} (\bibinfo{year}{2001}).

\bibitem[{\citenamefont{Giessibl et~al.}(2004)\citenamefont{Giessibl,
  Hembacher, Herz, Schiller, and Mannhart}}]{Giessibl:2004}
\bibinfo{author}{\bibfnamefont{F.~J.} \bibnamefont{Giessibl}},
  \bibinfo{author}{\bibfnamefont{S.}~\bibnamefont{Hembacher}},
  \bibinfo{author}{\bibfnamefont{M.}~\bibnamefont{Herz}},
  \bibinfo{author}{\bibfnamefont{C.}~\bibnamefont{Schiller}}, \bibnamefont{and}
  \bibinfo{author}{\bibfnamefont{J.}~\bibnamefont{Mannhart}},
  \bibinfo{journal}{Nanotechnology} \textbf{\bibinfo{volume}{15}},
  \bibinfo{pages}{S79} (\bibinfo{year}{2004}).

\bibitem[{\citenamefont{Israelachvili}(1991)}]{Israelachvili:1991}
\bibinfo{author}{\bibfnamefont{J.}~\bibnamefont{Israelachvili}},
  \emph{\bibinfo{title}{Intermolecular and Surface Forces, 2nd ed.}}
  (\bibinfo{publisher}{{}Academic Press, London}, \bibinfo{year}{1991}).

\bibitem[{\citenamefont{Albrecht et~al.}(1991)\citenamefont{Albrecht, Grutter,
  Horne, and Rugar}}]{Albrecht:1991}
\bibinfo{author}{\bibfnamefont{T.~R.} \bibnamefont{Albrecht}},
  \bibinfo{author}{\bibfnamefont{P.}~\bibnamefont{Grutter}},
  \bibinfo{author}{\bibfnamefont{H.~K.} \bibnamefont{Horne}}, \bibnamefont{and}
  \bibinfo{author}{\bibfnamefont{D.}~\bibnamefont{Rugar}}, \bibinfo{journal}{J.
  Appl. Phys.} \textbf{\bibinfo{volume}{69}}, \bibinfo{pages}{668}
  (\bibinfo{year}{1991}).

\bibitem[{\citenamefont{H\ölscher et~al.}(1999)\citenamefont{H\ölscher, Allers,
  Schwarz, Schwarz, and Wiesendanger}}]{Hoelscher:1999}
\bibinfo{author}{\bibfnamefont{H.}~\bibnamefont{H\ölscher}},
  \bibinfo{author}{\bibfnamefont{W.}~\bibnamefont{Allers}},
  \bibinfo{author}{\bibfnamefont{U.~D.} \bibnamefont{Schwarz}},
  \bibinfo{author}{\bibfnamefont{A.}~\bibnamefont{Schwarz}}, \bibnamefont{and}
  \bibinfo{author}{\bibfnamefont{R.}~\bibnamefont{Wiesendanger}},
  \bibinfo{journal}{{}Appl. Surf. Sci.} \textbf{\bibinfo{volume}{140}},
  \bibinfo{pages}{344} (\bibinfo{year}{1999}).

\bibitem[{\citenamefont{Guggisberg et~al.}(2000)\citenamefont{Guggisberg,
  Bammerlin, Baratoff, L\üthi, Loppacher, Battiston, L\ü, Bennewitz, Meyer, and
  G\üntherodt}}]{Guggisberg:2000b}
\bibinfo{author}{\bibfnamefont{M.}~\bibnamefont{Guggisberg}},
  \bibinfo{author}{\bibfnamefont{M.}~\bibnamefont{Bammerlin}},
  \bibinfo{author}{\bibfnamefont{A.}~\bibnamefont{Baratoff}},
  \bibinfo{author}{\bibfnamefont{R.}~\bibnamefont{L\üthi}},
  \bibinfo{author}{\bibfnamefont{C.}~\bibnamefont{Loppacher}},
  \bibinfo{author}{\bibfnamefont{F.}~\bibnamefont{Battiston}},
  \bibinfo{author}{\bibfnamefont{J.}~\bibnamefont{L\ü}},
  \bibinfo{author}{\bibfnamefont{R.}~\bibnamefont{Bennewitz}},
  \bibinfo{author}{\bibfnamefont{E.}~\bibnamefont{Meyer}}, \bibnamefont{and}
  \bibinfo{author}{\bibfnamefont{H.-J.} \bibnamefont{G\üntherodt}},
  \bibinfo{journal}{Surf. Sci.} \textbf{\bibinfo{volume}{461}},
  \bibinfo{pages}{255} (\bibinfo{year}{2000}).

\bibitem[{\citenamefont{Bennewitz et~al.}(2001)\citenamefont{Bennewitz, Sch\är,
  Barwich, Pfeiffer, Meyer, Krok, Such, Kolodzej, and
  Szymonski}}]{Bennewitz:2001}
\bibinfo{author}{\bibfnamefont{R.}~\bibnamefont{Bennewitz}},
  \bibinfo{author}{\bibfnamefont{S.}~\bibnamefont{Sch\är}},
  \bibinfo{author}{\bibfnamefont{V.}~\bibnamefont{Barwich}},
  \bibinfo{author}{\bibfnamefont{O.}~\bibnamefont{Pfeiffer}},
  \bibinfo{author}{\bibfnamefont{E.}~\bibnamefont{Meyer}},
  \bibinfo{author}{\bibfnamefont{F.}~\bibnamefont{Krok}},
  \bibinfo{author}{\bibfnamefont{B.}~\bibnamefont{Such}},
  \bibinfo{author}{\bibfnamefont{J.}~\bibnamefont{Kolodzej}}, \bibnamefont{and}
  \bibinfo{author}{\bibfnamefont{M.}~\bibnamefont{Szymonski}},
  \bibinfo{journal}{{}Surf. Sci.} \textbf{\bibinfo{volume}{474}},
  \bibinfo{pages}{L197} (\bibinfo{year}{2001}).

\bibitem[{\citenamefont{Allers et~al.}(2002)\citenamefont{Allers, Schwarz, and
  Schwarz}}]{Allers:2002bk}
\bibinfo{author}{\bibfnamefont{W.}~\bibnamefont{Allers}},
  \bibinfo{author}{\bibfnamefont{A.}~\bibnamefont{Schwarz}}, \bibnamefont{and}
  \bibinfo{author}{\bibfnamefont{U.~D.} \bibnamefont{Schwarz}}, in
  \emph{\bibinfo{booktitle}{Noncontact Atomic Force Microscopy}}, edited by
  \bibinfo{editor}{\bibfnamefont{S.}~\bibnamefont{Morita}},
  \bibinfo{editor}{\bibfnamefont{R.}~\bibnamefont{Wiesendanger}},
  \bibnamefont{and} \bibinfo{editor}{\bibfnamefont{E.}~\bibnamefont{Meyer}}
  (\bibinfo{publisher}{Springer Berlin Heidelberg New York},
  \bibinfo{year}{2002}), chap.~\bibinfo{chapter}{14}, pp.
  \bibinfo{pages}{233--256}.

\bibitem[{\citenamefont{Perez et~al.}(1998)\citenamefont{Perez, Stich, Payne,
  and Terakura}}]{Perez:1998}
\bibinfo{author}{\bibfnamefont{R.}~\bibnamefont{Perez}},
  \bibinfo{author}{\bibfnamefont{I.}~\bibnamefont{Stich}},
  \bibinfo{author}{\bibfnamefont{M.~C.} \bibnamefont{Payne}}, \bibnamefont{and}
  \bibinfo{author}{\bibfnamefont{K.}~\bibnamefont{Terakura}},
  \bibinfo{journal}{Phys. Rev. B} \textbf{\bibinfo{volume}{58}},
  \bibinfo{pages}{10835} (\bibinfo{year}{1998}).

\bibitem[{\citenamefont{Giessibl}(1992)}]{Giessibl:1992a}
\bibinfo{author}{\bibfnamefont{F.~J.} \bibnamefont{Giessibl}},
  \bibinfo{journal}{Phys. Rev. B (RC)} \textbf{\bibinfo{volume}{45}},
  \bibinfo{pages}{13815} (\bibinfo{year}{1992}).

\bibitem[{\citenamefont{Giessibl}(1997)}]{Giessibl:1997b}
\bibinfo{author}{\bibfnamefont{F.~J.} \bibnamefont{Giessibl}},
  \bibinfo{journal}{Phys. Rev. B} \textbf{\bibinfo{volume}{56}},
  \bibinfo{pages}{16010} (\bibinfo{year}{1997}).

\bibitem[{\citenamefont{Ueyama et~al.}(1998)\citenamefont{Ueyama, Sugawara, and
  Morita}}]{Ueyama:1998}
\bibinfo{author}{\bibfnamefont{H.}~\bibnamefont{Ueyama}},
  \bibinfo{author}{\bibfnamefont{Y.}~\bibnamefont{Sugawara}}, \bibnamefont{and}
  \bibinfo{author}{\bibfnamefont{S.}~\bibnamefont{Morita}},
  \bibinfo{journal}{{}Appl. Phys. A} \textbf{\bibinfo{volume}{66}},
  \bibinfo{pages}{S295} (\bibinfo{year}{1998}).

\bibitem[{\citenamefont{Giessibl and Trafas}(1994)}]{Giessibl:1994a}
\bibinfo{author}{\bibfnamefont{F.~J.} \bibnamefont{Giessibl}} \bibnamefont{and}
  \bibinfo{author}{\bibfnamefont{B.~M.} \bibnamefont{Trafas}},
  \bibinfo{journal}{Rev. Sci. Instrum.} \textbf{\bibinfo{volume}{65}},
  \bibinfo{pages}{1923} (\bibinfo{year}{1994}).

\bibitem[{\citenamefont{Giessibl}(2000)}]{Giessibl:2000b}
\bibinfo{author}{\bibfnamefont{F.~J.} \bibnamefont{Giessibl}},
  \bibinfo{journal}{{}Appl. Phys. Lett.} \textbf{\bibinfo{volume}{76}},
  \bibinfo{pages}{1470} (\bibinfo{year}{2000}).

\bibitem[{\citenamefont{Repp et~al.}(2004)\citenamefont{Repp, Meyer, Olsson,
  and Persson}}]{Repp:2004}
\bibinfo{author}{\bibfnamefont{J.}~\bibnamefont{Repp}},
  \bibinfo{author}{\bibfnamefont{G.}~\bibnamefont{Meyer}},
  \bibinfo{author}{\bibfnamefont{F.~E.} \bibnamefont{Olsson}},
  \bibnamefont{and} \bibinfo{author}{\bibfnamefont{M.}~\bibnamefont{Persson}},
  \bibinfo{journal}{{}Science} \textbf{\bibinfo{volume}{305}},
  \bibinfo{pages}{493} (\bibinfo{year}{2004}).

\end{thebibliography}

\end{document}